\documentclass[12pt,aps,prd,showpacs,nofootinbib]{revtex4}
\usepackage{amsfonts,amssymb,latexsym}

\newcommand{\bls}[1]{\renewcommand{\baselinestretch}{#1}}

\def\cm{\hspace*{1cm}}

\def\Jl#1#2{#1 {\bf #2},\ }

\def\ApJ#1 {\Jl{Astroph. J.}{#1}}
\def\CQG#1 {\Jl{Class. Quantum Grav.}{#1}}
\def\DAN#1 {\Jl{Dokl. AN SSSR}{#1}}
\def\GC#1 {\Jl{Grav. Cosmol.}{#1}}
\def\GRG#1 {\Jl{Gen. Rel. Grav.}{#1}}
\def\JETF#1 {\Jl{Zh. Eksp. Teor. Fiz.}{#1}}
\def\JETP#1 {\Jl{Sov. Phys. JETP}{#1}}
\def\JHEP#1 {\Jl{JHEP}{#1}}
\def\JMP#1 {\Jl{J. Math. Phys.}{#1}}
\def\NPB#1 {\Jl{Nucl. Phys.}{B\ #1}}
\def\NP#1 {\Jl{Nucl. Phys.}{#1}}
\def\PLA#1 {\Jl{Phys. Lett.}{#1A}}
\def\PLB#1 {\Jl{Phys. Lett.}{#1B}}
\def\PRD#1 {\Jl{Phys. Rev.}{D\ #1}}
\def\PRL#1 {\Jl{Phys. Rev. Lett.}{#1}}

\def\al{&\!}                            \def\lal{&&\!\! {}}
\def\eq{Eq.\,}                          \def\eqs{Eqs.\,}
\def\beq{\begin{equation}}              \def\eeq{\end{equation}}
\def\bear{\begin{eqnarray}}             \def\bearr{\begin{eqnarray} \lal}
\def\ear{\end{eqnarray}}                \def\earn{\nonumber \end{eqnarray}}
\def\nn{\nonumber\\ {}}                 
\def\nnn{\nonumber\\ \lal }             
                     \def\yyy{\\[5pt] \lal }
\def\eql{\al =\al}

\def\dst{\displaystyle}                 \def\tst{\textstyle}
\def\fracd#1#2{{\dst\frac{#1}{#2}}}     \def\fract#1#2{{\tst\frac{#1}{#2}}}
\def\Half{{\fracd{1}{2}}}               \def\half{{\fract{1}{2}}}

\def\e{{\,\rm e}}                       \def\d{\partial}

\def\sign{\mathop{\rm sign}\nolimits}   \def\diag{\mathop{\rm diag}\nolimits}
     \def\const{{\rm const}}
\def\eps{\varepsilon}

\def\DAL{\mathop{\square}\nolimits}
\newcommand{\vars}[1]{\left\{\begin{array}{ll}#1\end{array}\right.}

\def\og{\overline{g}}
\def\oR{\overline{R}}
\def\cG{{\cal G}}

\def\mn{_{\mu\nu}}                     \def\MN{^{\mu\nu}}
\def\mN{_\mu^\nu}

\def\ssph{static, spherically symmetric}

\def\bh{black hole}                    \def\bhs{black holes}
\def\wh{wormhole}                      \def\whs{wormholes}
\def\asflat{asymptotically flat}

\bls{1.00}

\begin{document}

\title{Spherical systems in models of nonlocally corrected gravity}

\author{K.A. Bronnikov}
\affiliation{Center for Gravitation and Fundamental Metrology, VNIIMS,
 46 Ozyornaya St., Moscow 119361, Russia;\\
 Institute of Gravitation and Cosmology, PFUR, 6 Miklukho-Maklaya St.,
 Moscow 117198, Russia} \email {kb20@yandex.ru}

\author{E. Elizalde}
\affiliation{Consejo Superior de Investigaciones Cient\'{\i}ficas
 ICE/CSIC-IEEC \\ Campus UAB, Facultat de Ci\`encies, Torre C5-Parell-2a pl,
 E-08193 Bellaterra (Barcelona) Spain}
\email {elizalde@ieec.uab.es, elizalde@math.mit.edu}

\begin{abstract}
  The properties of \ssph\ configurations are considered in the framework
  of two models of nonlocally corrected gravity, suggested in
  S. Deser and R. Woodard., Phys. Rev. Lett. {\bf 663}, 111301 (2007), and
  S. Capozziello et al., Phys. Lett. B {\bf 671}, 193--198 (2009).
  For the first case, where the Lagrangian of nonlocal origin represents a
  scalar-tensor theory with two massless scalars, an explicit condition is
  found under which both scalar fields are canonical (non-phantom). If this
  condition does not hold, one of the fields exhibits a phantom behavior.
  Scalar-vacuum configurations then behave in a manner known for
  scalar-tensor theories. In the second case, the Lagrangian of nonlocal
  origin exhibits a scalar field interacting with the Gauss-Bonnet (GB)
  invariant and contains an arbitrary scalar field potential.
  It is found that the GB term, in general, leads to violation of the
  well-known no-go theorems valid for minimally coupled scalar fields in
  general relativity. It is shown, however, that some configurations of
  interest are still forbidden --- whatever be the scalar field potential and
  the GB-scalar coupling function, namely, ``force-free'' wormholes
  (such that $g_{tt}= \const$) and \bhs\ with higher-order horizons.
\end{abstract}


\maketitle

\section{Introduction}

  It has been recently shown  (see \cite{nloc1,nloc2,nloc3} and
  references therein) that the dynamical Casimir effect manifests itself in
  the effective models of gravity and cosmology (owing, in particular, to the
  possible existence of compact extra dimensions) in the appearance of
  nonlocal contributions to the effective gravitational field Lagrangian.
  Cosmological consequences of such theories have been widely discussed
  (see, e.g., the same references above); less attention has been paid,
  however, to local configurations, such as stellar models, black holes,
  wormholes etc., whose existence and properties can crucially depend on
  nonlocal corrections to gravity and are very important in astrophysical
  observations. Moreover, some of these configurations can lead to
  interesting links between \bh\ physics and cosmology as, for instance, the
  ``black universes'' described in \cite{pha1, pha2}, which look like black
  holes from one of their two asymptotic regions and like an expanding de
  Sitter universe from the other.

  In this paper, we will study the possible impact of nonlocal gravity
  corrections on the existence and properties of the simplest local objects,
  namely, \ssph\ configurations, such as \bhs\ and \whs. To our knowledge,
  this is to date the first attempt to study wormholes in nonlocal gravity.
  Two particular models of nonlocal gravity will be considered, namely, those
  introduced in \cite{des07,nloc3}.

\section{Effective multiscalar-tensor theory: normal and phantom behavior}

  One version \cite{nloc2} of nonlocally corrected gravity in four dimensions
  is described by the Lagrangian
\beq                                                             \label{L0}
      L\sqrt{-g} = \sqrt{-g}\Big\{
             \Half R\Big[ 1+ f(\DAL^{-1} R) \Big] + L_m \Big\},
\eeq
  where $R$ is the Ricci scalar, $L_m$ is the Lagrangian of matter, and
  $\DAL$ the d'Alembertian operator. It has been shown \cite{nloc4} that
  this Lagrangian can be cast into a local form:
\beq                                                             \label{L1}
    L\sqrt{-g} = \sqrt{-g}\Big\{
     \Half \Big[ R\left(1 + f(\phi) -\xi\right) -\d_\mu \xi \d^\mu \phi\Big]
     		+ L_m \Big\},
\eeq
  where $\phi$ and $\xi$ are scalar fields. \eq (\ref{L1}) represents the
  Jordan frame of a scalar-tensor theory with two massless scalars. It
  should be noted that this kind of Lagrangians is a typical manifestation
  of the Casimir effect, since the nonlocal theory is a direct consequence of
  quantum field theory effects in the curved spacetime \cite{zeta}. We are
  thus here considering a direct application of a semiclassical description
  of the Casimir effect to the current cosmological epoch, which is
  interesting to remark.

  To be precise, the Lagrangian (\ref{L1}) leads to some extra solutions
  as compared to (\ref{L0}), as has been demonstrated \cite{kosh09} at least
  in the degenerate case $f = \const$ (see also \cite{pons1}).
  A reason is that the derivation of (\ref{L1}) involves substitution
  of the second-order constraint $\DAL \phi = R$ into the action,
  possibly introducing extra degrees of freedom.
  Thus some extra care is required with the initial conditions for
  the equations of motion due to (\ref{L1}) for all solutions to coincide
  with those due to (\ref{L0}). Actually this will not affect our results
  here, in our non-degenerate case, once we restrict to the ``mass shell"
  (the second-order constraint condition, in this case). On a more general
  setting of the correspondence problem between nonlocal theories and their
  local counterparts see, e.g., \cite{vernov09} and references therein.

  To make clear under what conditions the scalars in (\ref{L1}) have usual
  kinetic terms, and whether or not they contain a phantom degree of freedom,
  it is helpful to pass on to the Einstein-frame metric $\og\mn$ using the
  standard conformal mapping
\beq
      g\mn = \frac{1}{F} \og\mn,                               \label{eq2}
\eeq
  where
\beq                                                           \label{eq3}
	F = F(x^\mu) = 1 + f(\phi) -\xi
\eeq
  is the coefficient of $R$ in (\ref{L1}) representing the nonminimal
  coupling of the scalars $\phi$ and $\xi$ to gravity. We assume $F > 0$,
  to provide for a positive effective gravitational constant in the theory
  (\ref{L1}). This assumption is thus necessary for a meaningful theory and
  manifestly holds if $f$ and $\xi$ are small, that is, closely enough to the
  limit in which the theory (\ref{L0}) approaches general relativity.
  Moreover, the fact that under this assumption we afterwards arrive at the
  well-defined sigma-model Lagrangian (\ref{eq8}) with two scalar fields
  shows that $F > 0$ does not too strongly restrict the set of solutions.

  The transformation (\ref{eq2}) results in
\bear                                                       \label{eq4}
     L\sqrt{-g} = \Half \sqrt{\og} \biggl\{\oR - \frac{1}{F^2}
      \biggl[\frac{3}{2} (\d F)^2 + F \og\MN \d_\mu\phi \d_\nu\xi\biggr]
		+ \frac{2L_m}{F^2} \biggr\},
\ear
  where bars mark quantities obtained from or with the metric $\og\mn$, and
  $(\d F)^2 = \og\MN \d_\mu F \d_\nu F$.

  This Lagrangian describes a massless nonlinear sigma model with two scalars
  $\phi$ and $\xi$. It is a special case of nonlinear sigma models of the form
\beq                                                        \label{sigma}
    L = \Half R - h_{ab} \og\MN \phi^a \phi^b + \frac{L_m}{F^2},
\eeq
  where $h_{ab}$ are arbitrary functions of $n$ scalar fields $\phi^a$
  ($a, b = 1, 2, \ldots, n$). If the matrix $h_{ab}$ is positive-definite,
  the set of scalar fields is normal (non-phantom) in the sense that the
  kinetic energy is positive. To check that one can, as usual, diagonalize
  $h_{ab}$ algebraically, as is conventionally done for quadratic forms. It
  should be noted, however, that such a procedure will not, in general, lead
  to a valid Lagrangian in terms of the newly introduced scalar fields,
  because linear combinations of derivatives as, e.g., $A(\phi,\xi) \d_\mu
  \phi + B(\phi,\xi) \d_\mu \xi$, are not always integrable, and it can be
  quite hard to find an integrating factor.

  We therefore try to diagonalize the kinetic term in (\ref{eq4})
  by substituting
\beq                                                        \label{eq5}
      F = F(\phi, \eta), \cm \xi = 1 + f(\phi) - F,
\eeq
  where $\eta$ is a new field introduced instead of $\xi$; the second
  equality is just a rewriting of (\ref{eq3}). Then, the kinetic term in
  (\ref{eq4}) takes the form
\beq                                                        \label{eq6}
      \frac{3}{2F^2} \biggl\{
      	  (\d\phi)^2 \biggl[F_\phi^2 +\frac 23 F (f_\phi-F_\phi)\biggr]
	      + (\d\eta)^2 F_\eta^2
	      + 2 F_\eta \og\MN \d_\mu\phi\d_\nu\eta (F_\phi- F/3) \biggr\},
\eeq
  where the indices $\phi$ and $\eta$ denote $\d/\d \phi$ and $\d/\d \eta$,
  respectively. The expression (\ref{eq6}) is diagonal with respect to
  $\phi$ and $\eta$ under the conditions $F_\eta\ne 0$, $F_\phi = F/3$, and
  to satisfy them we choose simply
\beq
      F(\phi, \eta) = \eta \e^{\phi/3}.                       \label{eq7}
\eeq
  As a result, the Lagrangian (\ref{eq4}) reads
\beq                                                          \label{eq8}
    L = \Half R - \frac{3}{4} \frac{(\d \eta)^2}{\eta^2}
		- \frac{1}{12 F}(6f_\phi - F) (\d\phi)^2  + \frac{L_m}{F^2}.
\eeq

  We see that the theory is free from phantom fields if $6f_\phi > F$
  and that it does contain a phantom if $6f_\phi < F$. A more general version
  of \eq (\ref{eq8}) in $D$ dimensions can be found in \cite{koiv08a}, where
  the Newtonian limit and post-Newtonian corrections of the theory (\ref{L1})
  were considered.

  The properties of the theory are well illustrated by \ssph\ vacuum
  solutions similar to Schwarzschild's in general relativity. Let us
  take the Einstein-frame Lagrangian in the general form (\ref{sigma}).
  Assuming $L_m =0$ (vacuum), we can easily find the corresponding metric,
  whose properties depend on whether the matrix $h_{ab}$ is
  positive-definite or not.

  Indeed \cite{gc09}, if we write the general \ssph\ metric as
\beq                                                        \label{sph}
     ds^2 = -\e^{2\gamma(u)}dt^2 + \e^{2\alpha(u)}du^2
     	    + \e^{2\beta(u)} d\Omega^2,
\eeq
  where	$d\Omega^2 = (d\theta^2 + \sin^2\theta d\varphi^2)$ and $u$ is an
  arbitrary radial coordinate, and assume $\phi^a = \phi^a(u)$, the
  stress-energy tensor of the scalar fields has the form
\beq
    T\mN = h_{ab} (\phi^a)' (\phi^b)' \diag(1, -1, 1, 1),    \label{eq9}
\eeq
  that is, it has the same structure as for a single massless scalar field
  (here and henceforth the prime denotes $d/du$). Therefore, the metric has
  the same form as in this simple case: for a normal scalar it is the Fisher
  solution \cite{fish48}, for a phantom one it was first found by Bergmann
  and Leipnik \cite{berg57} and is sometimes called ``anti-Fisher'' (by
  analogy with anti-de Sitter). Let us reproduce it in the simplest joint
  form suggested in \cite{br73}.

  Choosing the harmonic radial coordinate $u$, such that $\alpha(u) =
  2\beta(u) + \gamma(u)$, we easily solve two combinations of the Einstein
  equations for the metric (\ref{sph}), namely, $R^0_0 =0$ (whence
  $\gamma'' = 0$) and $R^0_0 + R^2_2=0$ (whence $\beta'' + \gamma'' =
  \e^{2(\beta+\gamma)}$). As a result, the metric has the form
\beq                                                             \label{eq10}
     ds_E^2 = -\e^{-2mu} dt^2 + \frac{\e^{2mu}}{s^2(k,u)}
       		\biggr[\frac{du^2}{s^2(k,u)} + d\Omega^2\biggl],
\eeq
  where $k$ and $m$ are integration constants
  while the function $s(k,u)$ is defined as
\beq                                                             \label{eq11}
     s(k,u) = \vars     {
                    k^{-1}\sinh ku,  \ & k > 0, \\
                                  u, \ & k = 0, \\
                    k^{-1}\sin ku,   \ & k < 0.     }
\eeq
  In addition, the ${1\choose 1}$ Einstein equation gives
\bearr                                                          \label{eq12}
     k^2 \sign k = m^2 + C_\phi,
\yyy
     C_\phi = h_{ab} (\phi^a)' (\phi^b)' = \const.            \label{eq13}
\ear
  The scalar field equations read
\beq                                                           \label{eq14}
    2 \Big[ h_{ab}(\phi^b)'\Big]'
    		+ \frac{\d h_{bc}}{\d\phi_a} (\phi^b)' (\phi^c)' =0
\eeq
  and obviously cannot be solved in a general form, but (\ref{eq13}) is their
  first integral. The metric (\ref{eq10}) is defined (without loss of
  generality) for $u >0$, it is flat at spatial infinity $u = 0$, and $m$ has
  the meaning of a Schwarzschild mass in proper units. Its properties
  crucially depend on the sign of $k$, which in turn depends on $C_{\phi}$,
  hence on the nature of the matrix $h_{ab}$.

  If $h_{ab}$ is positive-definite, we have $C_\phi > 0$ for all nontrivial
  scalar field configurations and obtain the Fisher metric: in this case
  $k > 0$ and the substitution $\e^{-2ku} = 1 - 2k/r \equiv P(r)$ converts
  (\ref{eq10}) into
\beq                                                            \label{eq15}
     ds_E^2 = P(r)^{a} dt^2 - P(r)^{-a} dr^2
                                 - P(r)^{1 - a} r^2 d\Omega^2,
\eeq
  where $a = m/k = (1 - C_\phi/k^2)^{1/2} < 1$ (we assume $m>0$). The
  solution is defined for $r > 2k$, and $r = 2k$ is a central singularity.
  The Schwarzschild metric is restored for $C_\phi=0$, $a=1$.
  If $h_{ab}$ is not positive-definite, $C_\phi$ can have any sign. Thus, for
  some nontrivial scalar field configurations we may have $C_\phi=0$, hence
  the Schwarzschild metric. For others, there can be $C_\phi < 0$ which
  correspond to ``anti-Fisher'' phantom-field metrics. In addition to
  singular metrics, they include (in case $k>0$) the so-called cold black
  holes with horizons of infinite area (see \cite{cold08} and references
  therein). In case $k < 0$, the substitution  $|k| u = \cot^{-1} (r/|k|)$
  converts the metric (\ref{eq10}) into
\beq                                                           \label{eq16}
    ds_E^2 =\e^{-2mu} dt^2 - \e^{2mu} [d r^2 - (k^2 + r^2) d\Omega^2],
\eeq
  which describes a traversable wormhole with different signs of mass at its
  two flat asymptotics. More details about the anti-Fisher metric can be
  found in \cite{cold08, gc09}.

  As to the theory (\ref{L1}), we can assert that its vacuum \ssph\ solution
  in the Einstein frame is characterized by the Fisher metric in case
  $6\, df/d\phi > F$ and contains families with both Fisher and anti-Fisher
  metrics in case $6\, df/d\phi < F$.

  The Jordan-frame metric is obtained from (\ref{eq10}) by the transformation
  (\ref{eq2}). To perform it we must know the function $F$, which is hard to
  find in a general form, due to the difficulty of solving the
  field equations (\ref{eq14}). But one can assert that if $F$ is everywhere
  positive and finite (including its limiting values at infinity and
  the other extreme of the $u$ range), then the mapping (\ref{eq2})
  transforms regular points to regular points, a flat infinity to a flat
  infinity, and a singularity to a singularity. Therefore, the main
  qualitative features of the (anti-)Fisher metric, e.g., the existence of
  wormholes, are preserved by the Jordan-frame metric as well. Also, since
  (anti-)Fisher metrics do not contain horizons of finite area, such horizons
  will also be absent in the Jordan frame.  Regarding cosmological
  applications of this theory, we can mention that cosmologies with two
  scalars, one of them being phantom, and with two canonical scalars but with
  a nontrivial potential, have been discussed in \cite{cosm1,cosm2}; see also
  \cite{koiv08b}.

\section{Scalar fields interacting with the Gauss-Bonnet invariant}

  The initial Lagrangian can contain, in addition to
  (\ref{L1}), terms with the Gauss-Bonnet invariant \cite{nloc3}
\[
    \cG = R^2 - 4 R\mn R\MN + R_{\mu\nu\rho\sigma}R^{\mu\nu\rho\sigma},
\]
  and functions of $\DAL^{-1}$, so that its local form can read, e.g.,
\beq                                                       \label{eq17}
    L\sqrt{-g} = \sqrt{-g}\Big\{
     \Half \Big[ R[1 + f(\phi) -\xi] -\d_\mu \xi \d^\mu \phi\Big]
     - V(\phi) + h(\phi){\cal G} + L_m \Big\},
\eeq
  where $h(\phi)$ is a function specified by some underlying theory.
  Anyhow, being quadratic in the curvature, this term can play a significant
  role only at sufficiently large curvatures, most probably (if $h$ is not
  too large and not too rapidly changing) close to the Planck level. Thus,
  the above results are also valid for the Lagrangian (\ref{eq17}) at
  reasonably small curvatures. In particular, the nonsingular metric
  (\ref{eq16}) should remain a solution in the whole space in case $V\equiv
  0$ at moderate parameter values; however, at strong curvatures and, in
  particular, near the singularities, addition of $h(\phi){\cal G}$ can
  drastically change the geometry, and such cases deserve a further study.

  Here we restrict ourselves to a somewhat more special effective Lagrangian
  of the above type, containing a single scalar field minimally coupled to
  the curvature, namely,
\beq                                                          \label{L2}
    L\sqrt{-g} = \frac{\sqrt{-g}}{2\kappa^2}\Big\{
 		R - \eps g\MN \d_\mu\phi \d_\nu\phi -V(\phi)
		+ h(\phi) \cG \Big\} + \sqrt{-g} L_m,
\eeq
  where the coefficient $\eps =\pm 1$ distinguishes normal ($\eps=1$) and
  phantom ($\eps=-1$) scalar fields, $V(\phi)$ is a potential, and $\cG$ is
  the Gauss-Bonnet (GB) invariant. The theory (\ref{L2}) may be called
  generalized dilatonic GB gravity, its special cases have been widely
  discussed in the context of a low-energy limit of string theory,
  see \cite{string} for a recent discussion.

  For the general form (\ref{eq10}) of the \ssph\ metric, with an arbitrary
  radial coordinate $u$, the GB invariant is calculated to be (the prime
  denotes $d/du$)
\bear                                                         \label{GB-gen}
      \cG = \frac{8 F'(u)}{r^2 \e^{\alpha+\gamma}}, \cm
      	F(u):= \e^{-\alpha+\gamma}\gamma' (\e^{-2\alpha}r'{}^2 -1),
\ear
  in agreement with the fact that a pure GB term in the Lagrangian makes a
  full divergence and does not contribute to the field equations. In the
  Lagrangian (\ref{L2}) the invariant $\cG$ appears with the factor $h(\phi)$
  and therefore does contribute to the scalar and gravitational field
  equations which can be written in the following form:
\bearr                                                        \label{e-phi}
       	\eps\Big(r^2\e^{\gamma - \alpha} \phi'\Big)'
		- r^2 \e^{\alpha+\gamma} V_\phi + 4h_\phi F' = 0,
\yyy                                                          \label{00}
	- G^t_t = \frac{\eps}{2} \e^{-2\alpha} \phi'^2 + V
      	        	- \frac{4}{r^2}
      		\Big[\e^{-\alpha}h'\big(\e^{-2\alpha} r'^2 -1\big)\Big]',
\yyy                                                          \label{11}
	- G^r_r = -\frac{\eps}{2} \e^{-2\alpha} \phi'^2 + V
      			+ \frac{4}{r^2}
      		\e^{-2\alpha}h'\gamma'\big(1 - 3\e^{-2\alpha} r'^2),
\yyy                                                          \label{22}
	- G^\theta_\theta = \frac{\eps}{2} \e^{-2\alpha} \phi'^2 + V
      			- \frac{4}{r}
      	        \e^{-\alpha-\gamma}\Big(\e^{\gamma-3\alpha}h'r'\gamma'\Big)',
\ear
  where the subscript $\phi$ denotes derivatives $d/d\phi$ and $G\mN$ are
  components of the Einstein tensor, $G\mN = R\mN - \half \delta\mN R$.

  It is straightforward to check that the last terms in (\ref{00})--(\ref{22})
  form a conservative tensor $T\mN(\cG)$, so that $\nabla_\nu T\mN(\cG) =0$.
  Therefore, as is usually the case with \ssph\ systems with scalar fields,
  the scalar equation (\ref{e-phi}) is a consequence of the Einstein
  equations (\ref{00})--(\ref{22}). With the functions $h(\phi)$ and $V(\phi)$
  specified and a coordinate (gauge) condition chosen,
  (\ref{e-phi})--(\ref{22}) is a well determined set of equations for the
  field $\phi(u)$ and two metric functions.

\subsection*{Weak fields}

  Since $\cG$ is quadratic in the curvature, it can play a significant role
  only at sufficiently large curvatures, most probably (if $h(\phi)$ is not
  too large and not too rapidly changing) close to the Planck level.
  In particular, if we consider \ssph\ configurations governed by the
  Lagrangian (\ref{L2}) and assume that the curvatures are reasonably small
  (which in most cases is true, except in an immediate neighborhood of
  singularities), then the general properties of such configurations, known
  for (\ref{L2}) without $\cG$ (see, e.g., \cite{vac1, vac2, pha1}) or,
  equivalently, with $h(\phi) =\const$, remain valid in the presence of $\cG$.

  Let us recall some of these properties, by writing the metric (\ref{eq10})
  with the aid of the so-called quasi-global radial coordinate $u$ (that is,
  taking the gauge condition $\alpha+\gamma = 0$ and denoting
  $\e^{2\gamma} = A(u)$)  and assuming $\phi=\phi(u)$:
\beq                                                          \label{ds-A}
      ds^2 = -A(u) dt^2 + \frac{du^2}{A(u)} + r^2(u)
  	      		       (d\theta^2 + \sin^2\theta d\varphi^2).
\eeq
 Properties of interest are then given by the theorems:
\begin{enumerate}    \itemsep 0pt
\item
     If $\eps = +1$, the function $r(u)$ cannot have a regular minimum,
     whatever be the potential $V(\phi)$. In other words, wormhole throats
     are impossible, to say nothing of \whs\ as global entities \cite{vac1}.
\item
     No-hair theorem \cite{ad-pear}. Suppose $\eps=+1$ and $V \geq 0$. Then
     the only asymptotically flat \bh\ solution to the field equations in the
     range $(u_h, \infty)$, where $u=u_h$ is the event horizon, comprises the
     Schwarzschild metric, $\phi = \const$ and $V \equiv 0$.
\item
     An asymptotically flat solution with a regular center (i.e., a
     particle-like, or star-like solution) is impossible if $\eps = +1$ and
     $V(\phi) \geq 0$, or if $\eps =-1$ and $V(\phi) \leq 0$ \cite{vac2}.
\item
     Global structure theorem \cite{vac1}. The function $B(u) = A(u)/r^2$
     cannot have a regular minimum. It follows that the system can contain no
     more then two Killing horizons, which are described as regular zeros of
     $A(u)$, and then the horizons are simple and bound a static region
     ($A > 0$).  A double horizon is possible, but it then separates two
     nonstatic (T) regions, and a static region is absent.
\end{enumerate}
     Theorem 4 is the most universal: it holds for any $\eps$ and $V(\phi)$
     and does not depend on assumptions about the asymptotic behavior;
     however, for an \asflat\ configuration it implies that there can be no
     more than one horizon. Thus the whole set of possible kinds of global
     causal structure is the same as in the Schwarzschild-de Sitter solution,
     despite the existence of a scalar field.

  As mentioned above, nonsingular solutions for (\ref{L2}) without $\cG$,
  such as those describing \whs\ and regular \bhs\ which can emerge without
  violating the no-go theorems, are likely to satisfy, approximately, the full
  equations due to (\ref{L2}), if the corresponding curvatures are very small
  as compared to the Planck one.

  In the full theory (\ref{L2}) the above theorems are no longer valid. This
  can be checked directly, using the explicit expression of $\cG$ for \ssph\
  metrics. In particular, the so-called dilatonic GB \bh\ solutions with
  nontrivial scalar fields and $V \equiv 0$ are well known (see, e.g.,
  \cite{GBBH1,GBBH2,GBBH3,GBBH4} and references therein), so that the no-hair
  theorem does not hold. We will show, however, that some general
  restrictions can be obtained for any choice of $h(\phi)$ and $V(\phi)$.

\subsection*{Possible \whs}

  Let us begin with possible \wh\ solutions and use the form (\ref{ds-A}) of
  the metric. The expression (\ref{GB-gen}) then simplifies to give
\beq                                                          \label{GB-A}
       \cG = \frac{F'}{r^2}, \cm F(u) := A'(Ar'{}^2 -1),
\eeq
  where, as before, the prime stands for $d/du$. The difference between \eqs
  (\ref{00}) and (\ref{11}) reads
\beq                                                           \label{r''}
     2r r'' = -\eps r^2 \phi'{}^2 + 4h'' (Ar'{}^2-1) + 8A r'r''h',
\eeq
  while the difference between (\ref{00}) and (\ref{22}) can be written in
  the form
\beq                                                        \label{B''}
     A'' r^2 - (r^2)''A \equiv
     (r^4 B')' = -2 - 8\sqrt{A} \Big[\sqrt{A}(Ar'{}^2-1)h'\Big]'
		    + 4r \Big [ AA' r'h'\Big]',
\eeq
  where $B(u) = A/r^2$.

  With a phantom field $\phi$ ($\eps = -1$), \whs\ manifestly do exist, and
  the simplest example of such an exact solution is given by the Ellis \wh\
  \cite{ellis} for which
\bearr                                                        \label{el-wh}
    A(u) \equiv 1, \qquad r^2(u) = u^2 + k^2, \qquad k = \const >0,
\nnn
    \phi (u) = \sqrt{2}\arctan (u/k), \qquad V(\phi) \equiv 0,
\nnn
     h(\phi) = h_0 + h_1 (u^3 + 3k^2 u),\qquad u = k \tan (\phi/\sqrt{2}),
\ear
  where $h_0$ and $h_1$ are integration constants. The metric is a special
  case ($m=0$) of (\ref{eq16}) (with a changed notation, $r\mapsto u$). In
  case $h_1 = 0$ we have $h = \const$, thus returning to the usual
  Einstein-scalar equations leading to the anti-Fisher solution discussed
  above, but in the general case $h_1\ne 0$ (\ref{el-wh}) is an exact
  solution of the full theory (\ref{L2}).

  It is of interest whether or not \wh\ solutions can exist for $\eps = +1$.
  If $h = \const$ (i.e., $\cG$ does not contribute to the field equations),
  then in this case $r'' \leq 0$, and $r(u)$ cannot have a minimum, i.e.,
  Theorem 1 holds. But in the general case minima of $r$ are not excluded. At
  an extremum of $r$, where $r'=0$, we have from (\ref{r''})
\beq                                                          \label{min}
     2r r'' = -\eps r^2 \phi'{}^2 - 4h'',
\eeq
  and, with a properly chosen function $h(\phi)$, the second term can lead to
  $r'' >0$ even for $\eps=+1$.

  This does not mean, however, that such objects necessarily exist.
  To illustrate it, consider a simplified system with $A(u) \equiv 1$,
  i.e., spaces without a gravitational force acting on bodies at rest.
  The system as a whole remains nontrivial, since \eq (\ref{00}) still
  contains a contribution from $h(\phi)$. However, \eqs (\ref{11}) and
  (\ref{22}) do not contain $h$ (because now $\gamma' = A'/(2A) \equiv 0$ and,
  thus, the terms with $h$ are eliminated), and their difference reads
\beq
       \eps\phi'^2 = \frac{1}{r^2}(-1 + r'^2 - rr'').
\eeq
  It follows that a minimum of $r$, i.e., a \wh\ throat, where $r'=0$ and
  $r'' > 0$, can only occur with $\eps = -1$, just as was the case without a
  GB term. It should be stressed that this result does not depend on the
  choice of the coupling function $h(\phi)$ and the potential $V(\phi)$.

  In the general case $A(u) \ne \const$, \wh\ solutions are not excluded but
  it seems that, if any, they cannot be of great physical interest for the
  following reasons. Considering for certainty symmetric \whs\ for which
  simultaneously $r'(u)$ and $A'(u)$ vanish on the throat, from \eq
  (\ref{B''}) we obtain
\[
      r^2 A'' = -2 + 2Arr'' + 8A h'' \qquad \mbox{on the throat.}
\]
  Combined with (\ref{min}), this leads for $\eps = +1$ to the following
  inequality:
\beq                                                           \label{max}
     r^2 A'' < -2 - 2Arr''
\eeq
  on the throat where, by definition, $r'' > 0$. It means that $A(u)$  has
  quite a strong maximum there, i.e., the throat gravitationally repels test
  bodies. And, in any case, it is clear that in such \whs\ the curvature
  components should be of sub-Planckian order near the throat, hence either
  its size or the magnitude of tidal forces (or maybe both) make these
  \whs\ actually non-traversable for any macroscopic bodies.

\subsection*{Possible horizons}

  Let us return to \eq (\ref{B''}).
  In the absence of the Gauss-Bonnet invariant $\cG$, and consequently terms
  containing $h$, we have there only $-2$ on the r.h.s., which prevents a
  minimum of $B(u)$ and thus leads to Theorem 4. Indeed, a horizon is a zero
  of $A(u)$ at finite $r$, hence a zero of $B(u)$, and the absence of its
  minima leaves only a restricted list of possible allocations of such zeros.

  The terms with $h(\phi)$ substantially change the situation, and one may
  expect configurations with more complex global structures. However, it can
  be shown that double horizons, like that of the extremal
  Reissner-Nordstr\"om \bh\, as well as horizons of orders higher than 2,
  cannot exist in our system, whatever be the choice of $h(\phi)$ and
  $V(\phi)$.

  Indeed, let there be a horizon at $u=0$ and consider near-horizon
  Taylor expansions of all functions involved in \eq (\ref{B''}), namely,
\bear                                                           \label{hor}
       A(u) \eql A_1 u + \half A_2 u^2 + \ldots,
\nn
       r(u) \eql r_0 + r_1 u + \half r_2 u^2 + \ldots,
\nn
       h(\phi) \eql h(\phi(u)) = h_0 + h_1 u + \half h_2 u^2 + \ldots,
\ear
  and substitute them into (\ref{B''}). Note that the expansion (\ref{hor})
  in terms of the quasiglobal coordinate $u$ is a necessary feature of
  horizons: attempts to find other kinds of near-horizon behavior of the
  metric lead to singularities \cite{we08}.

  If $A_1 \ne 0$, i.e., there is a simple horizon, at order $O(1)$
  we obtain a relation between the constants,
\[
	A_2 r_0^2 - 4A_1 h_1 - 4r_0 r_1 A_1^2 h_1 +2 =0,
\]
  and the next orders involve coefficients from other terms of the expansion.
  Simple horizons are thus possible, in accordance with the known examples
  \cite{GBBH1,GBBH2,GBBH4}.

  If $A_1 =0$ and $A_2 \ne 0$, i.e., there is a double horizon,
  \eq (\ref{hor}) at order $O(1)$ reads
\[
	r_0^2 A_2 + 2 =0,
\]
  whence $A_2 < 0$, i.e., a double horizon is possible but $A < 0$ in its
  neighborhood. Thus such a horizon can separate two T regions but not two R
  regions, a horizon like the extreme Reissner-Nordstr\"om one is impossible.
  The situation is, in this respect, the same as the one described by Theorem
  4.

  If the first nonzero coefficient $A_n$ is with $n>2$ (which means a
  higher-order or ultra-extremal horizon), the only term of order $O(1)$
  in (\ref{hor}) is $-2$, making the equation inconsistent. Such horizons are
  thus impossible.

\section{Concluding remarks}

  For the Lagrangian (\ref{L1}) of nonlocal origin, representing the Jordan
  frame of a scalar-tensor theory with two massless scalars, we have found
  an explicit condition under which both scalar fields are canonical
  (non-phantom). If this condition does not hold, one of the fields
  exhibits a phantom behavior. The properties of the corresponding
  scalar-vacuum \ssph\ configurations are well known, and we have briefly
  described them here for clarity. They include geometries with naked
  singularities and, in the phantom case, traversable \whs.

  For the Lagrangian (\ref{L2}) of nonlocal origin, containing a scalar
  field interacting with the Gauss-Bonnet invariant and a nonzero
  scalar field potential, we have found that the Gauss-Bonnet term,
  in general, leads to violation of the well-known no-go theorems valid
  for minimally coupled scalar fields in general relativity. This means
  that many configurations of interest, forbidden by these theorems,
  can appear, but only if curvatures are sufficiently large, approaching
  the Planck value.

  We have shown, however, that some configurations of interest are still
  forbidden even in the full theory, whatever be the scalar field potential
  $V(\phi)$ and the GB-scalar coupling function $h(\phi)$. Among these
  configurations there are ``force-free'' wormholes with a normal ($\eps=+1$)
  scalar field (i.e., \whs\ with $g_{tt} = \const$) and \bhs\ with
  higher-order (extremal or ultraextremal) horizons. The fact that these
  considerations are based on the local form of the Lagrangian---which may
  lead in principle to extra solutions as compared to the nonlocal
  form (\ref{L1})---poses no problem, in the end. To wit, in the more
  standard theory the class of solutions is wider but each of them is
  given by a single-valued (ordinary) function, while in the nonlocal
  theory the class of solutions (which is more restricted) contains
  multi-valued functions. Eventually, this does not affect our
  ``on shell" results (once the second-order constraint condition
  relating the two formulations is imposed).

  According to \cite{GBBH3}, double (extremal) horizons are possible in the
  theory (\ref{L2}) in configurations with an electric charge. They are found
  and discussed in the special case $h(\phi)\sim \e^{a\phi}$, $a = \const$,
  $V(\phi) \equiv 0$. We have shown that such horizons cannot exist in
  vacuum (without an electromagnetic field) in the general case of the theory
  (\ref{L2}), for any $h(\phi)$ and any potential $V(\phi)$. Such extremal
  \bh\ solutions, in general, separate configurations without horizons (e.g.,
  with naked singularities) from \bhs\ with two simple horizons. This
  happens, for instance, in the Reissner-Nordstr\"om metric and in the family
  of solutions of Einstein gravity coupled to nonlinear electrodynamics (see,
  e.g., \cite{br-ned}). In the present case, the non-existence of double
  horizons probably means that \bhs\ with two simple horizons and a
  Reissner-Nordstr\"om-like causal structure are also absent in the general
  case of the theory (\ref{L2}). But this is a conjecture yet to be proved.

\subsection* {Acknowledgments}

  We thank Sergei Odintsov for many helpful discussions and Tomi Koivisto for
  useful comments to the first version of the paper. KB acknowledges kind
  hospitality at ICE/CSIC-IEEC and partial financial support from the ESF
  Exchange Grant 2040 within the activity ``New Trends and Applications of
  the Casimir Effect'', and from the NPK MU grant at PFUR.  Part of EE's
  research was performed while on leave at Department of Physics and
  Astronomy, Dartmouth College, 6127 Wilder Laboratory, Hanover, NH 03755,
  USA. This work was partly supported by MEC (Spain), project FIS2006-02842,
  and by AGAUR (Generalitat de Catalunya), contract 2009SGR-994 and grant
  DGR2009BE-1-00132.

\end{document}